# Extremely broadband topological surface states in a photonic topological metamaterials

Minkyung Kim[1], Dasol Lee[1], Wenlong Gao[2], Taewoo Ha[3,4], Teun-Teun Kim[3,4], Shuang Zhang[2] and Junsuk Rho[1,5]*

[1]Department of Mechanical Engineering, Pohang University of Science and Technology (POSTECH), Pohang 37673, Republic of Korea

[2]School of Physics and Astronomy, University of Birmingham, Birmingham B15 2TT, United Kingdom

[3]Center for Integrated Nanostructure Physics (CINAP), Institute for Basic Science (IBS), Suwon 16419, Republic of Korea

[4]Sungkyunkwan University, Suwon 16419, Republic of Korea

[5]Department of Chemical Engineering, Pohang University of Science and Technology (POSTECH), Pohang 37673, Republic of Korea

E-mail: jsrho@postech.ac.kr

**Abstract**

Metamaterials, artificially engineered materials consisting of subwavelength unit cell, have shown potentials in light manipulation with their extraordinary optical properties. Especially, topological metamaterials possessing topologically protected surface states enable extremely robust control of light. Here, we demonstrate extremely broadband topological phase in a photonic topological metamaterials with double helix structure. In particular topological surface states are observed for all the frequencies below a certain cut-off, originating from a double Weyl point at zero frequency. The extreme bandwidth and robustness of the photonic topological metamaterial are beneficial for practical applications such as one-way waveguide and photonic integrated systems but also advantageous in design and fabrication since the only necessary condition is to satisfy the effective hyperbolic and chiral properties, without entailing strict periodic arrangement.

Recent discovery of Weyl fermion in certain crystals, which had been only previously predicted in quantum field theory [1, 2], have opened a new realm of condensed matter physics [3]. The photonic analogy of Weyl degeneracies has been soon implemented in photonics [4-8]. Besides Weyl points, other topological states, such as quantum Hall effect and quantum spin Hall effect have been also predicted and observed in photonics [9-18]. During the last several years, photonic topological materials have developed rapidly, revealing many exotic phenomena including photonic topological surface states, chiral anomalies and Weyl degeneracies. Weyl points, monopoles of Berry curvature in the momentum space with integer topological charges, are manifestations of the photonic Weyl systems, which are endowed with topological surface states, or photonic Fermi arcs that connect Weyl points with the opposite topological charges.

In the field of topological photonics, most of the studies have been performed in the regime of photonic crystal where periodicity is comparable to the wavelength. There also a few works based on metamaterials, artificially structured optical materials composed of unit cells much smaller than the wavelength [4, 5, 19]. The mechanism of the metamaterial-based approaches is fundamentally different from photonic crystal-based ones, relying on the combination of two distinct properties: hyperbolicity and chirality. Existence of topologically protected surface states and photonic Weyl points in a medium with hyperbolic and chiral responses have been theoretically predicted [4, 19] and experimentally demonstrated [5] recently. Since the topological surface states are formed between the Weyl point pairs, bandwidth of the surface states is related to the frequencies of the Weyl points. Here, we present a photonic topological metamaterial with a double Weyl point at zero frequency, generating extremely broadband topological surface states. The zero frequency Weyl point sets the lower limit of the bandwidth, so that the topological surface states exist at all frequencies below a certain frequency carrying the other Weyl point pair. Focused on three lowest bands, there exist six pairs of Weyl points, among which two have multiple topological charges. Chern numbers directly calculated from the simulated eigenmodes support our result. The presence of topological surface states across a very broad bandwidth may lead to potential applications in topologically protected integrated photonic circuits.

**Results**

**Bulk states dispersion of the topological metamaterial**

A typical hyerpbolic metamaterial consists of an array of metallic wire embedded in a dielectric medium with deep subwavelength spacings [20, 21]. This highly anisotropic geometry makes the wire system to have opposite signs of permittivity along the wires and perpendicular to the wires. However, straight wire array does not possess no chirality due to the mirror symmetry. To simultaneously achieve hyperbolic properties and chirality in a single structure, we introduce long metallic helices [22, 23]. Therefore, one can expect that metallic helix structure has both hyperbolic and chiral properties (Figure 1a), thereby possessing topologically non-trivial optical properties. Indeed, by adding a twist to the wire structure to break the inversion symmetry, the nodal line node is transformed into Weyl points [24].

A right-handed double helix structure is illustrated in Figure 1b. The retrieved effective parameters of the double helix structure contain opposite signs of permittivity and non-zero chirality, satisfying both hyperbolic and chiral conditions (details can be found in supplementary S1). We would like to note that while we employ periodically arranged structure to simulate band structures, effective hyperbolicity and chirality are the only necessary condition to achieve topological phase in the metamaterial regime whereas lattice system with exact periodic arrangement is essential in photonic crystals.

Geometric parameters of the double helix structure are set as the following: periodicity along x- and y-axis $a = 10$ mm, pitch $p = 10$ mm, helix radius $R = 3$ mm and fiber radius $r = 1$ mm. Figure 1b shows two unit cells aligned along z-axis for clear picture. Note that periodicity along z-axis is half of the pitch because the unit cell has screw symmetry; in other words, it is invariant under the $\pi/2$ rotation along z-axis followed by $p/4$ translation along z-axis. Considering the periodicity along z-axis, dispersion has $C_4$ symmetry along $k_z$-axis in a momentum space. The unit cell is designed to work at microwave regime for the ease of fabrication and near-field imaging measurement, but by expanding or shrinking the size of unit cell, the topologically non-trivial system can operate at any wavelength from visible to microwave range (See supplementary S6).

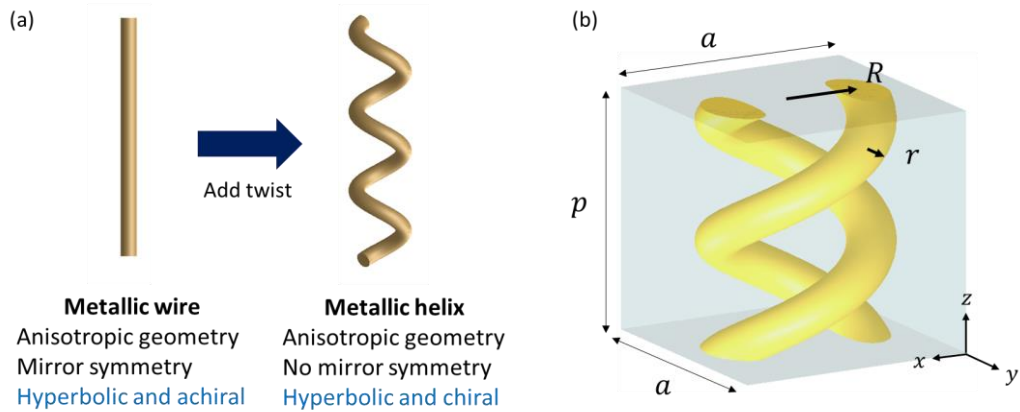

Figure 1. (a) Schematics for the design rule. (b) Two unit cells aligned along z-axis. The geometry parameters are given as: $a = 10$ mm, $p = 10$ mm, $R = 3$ mm and $r = 1$ mm.

We perform wave vector eigenvalue simulation described in [25, 26], and also use eigenfrequency solver in COMSOL Multiphysics to obtain bulk dispersion and equifrequency curve of the real structure. Perfect electric conductor condition is used for metallic parts and refractive index of the surrounding dielectric medium is set as 2. Band structures and equifrequency curves of the double helix structure are shown in Figure 2. Wave vectors are normalized by $\pi/a$ for $k_x$ and $k_y$ and $2\pi/p$ for $k_z$.

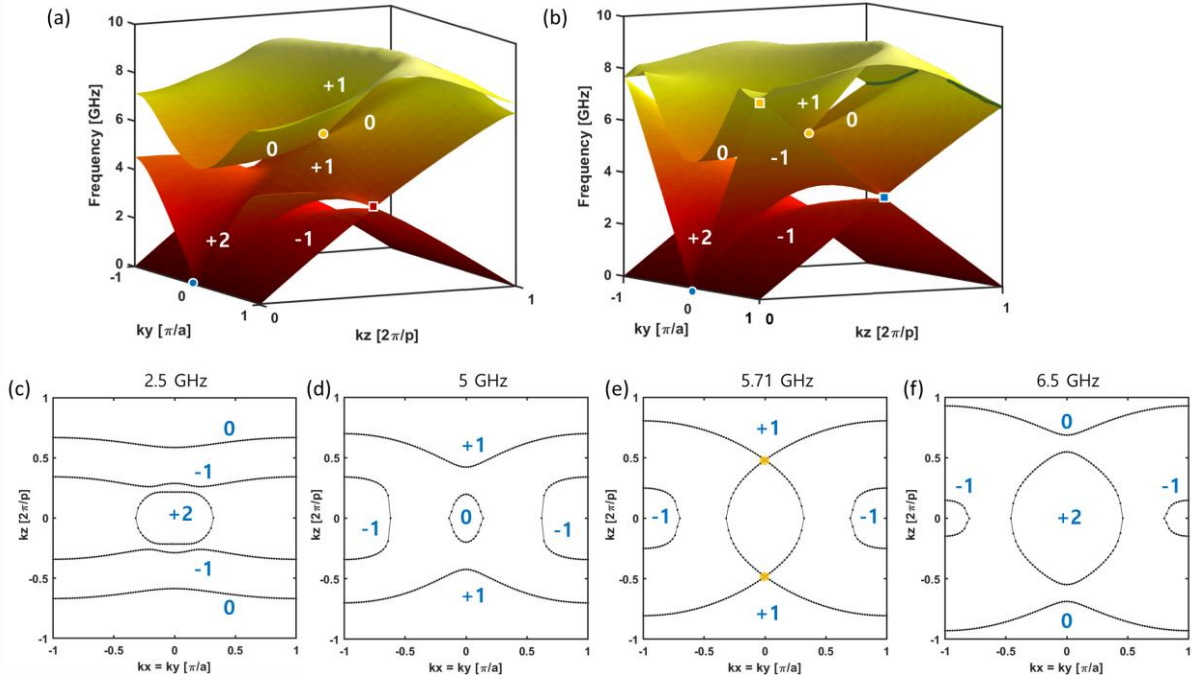

Figure 2. (a) and (b) Band structure of (a) $k_x = 0$ and (b) $k_x = k_y$ and Chern numbers. (c-f) Equifrequency curves along a line $k_x = k_y$ and Chern numbers at (c) 2.5 GHz, (d) 5 GHz, (e) 5.71 GHz and (f) 6.5 GHz. Circle and square points with the same color represent distinct Weyl points forming a pair.

Here we focus on the lowest three bands. The bands have several point and line degeneracies which are presented as blue, yellow and red points and green lines. Circle and square points with the same color represent distinct point degeneracies forming a pair. Chern numbers are calculated based on Wilson loop method by using spatially averaged fields instead of the effective Hamiltonian model (See supplementary S2). The sign of the Chern number is determined by the helical direction; The sign is flipped for the opposite handedness (left-handed double helix structure). The line degeneracy denoted as green line has no topological charge. On the other hand, the point degeneracies are all Weyl points with non-zero topological charges. Among them, the pair labelled as yellow circles is type 2 Weyl points and the others are type 1 Weyl points. Interestingly, two Weyl points, one at the Brillouin zone center (blue circle) and the other at the zone boundary (yellow square), are doubly charged and have two partners of single charge of the opposite sign. The double or even multiple Weyl point has been predicted and reported in condensed matter physics [27-30] and also in photonics [31, 32]. They have linear dispersion around the crossing point only in one reciprocal direction while nonlinear dispersion along the two other directions with multiple charges, quadratic for double and cubic for triple, as opposed to cone-shape dispersion near Weyl points of single topological charge [27, 28]. Nonlinear and linear dispersion around the Weyl point marked as yellow square can be found in Supplementary S3. The doubly charged Weyl points are protected by $C_4$ or $C_6$ point group symmetries [27]. As such, by breaking the $C_4$ symmetry through designing different helix radius along x- and y-axis, the double Weyl point is split into two type 2 Weyl points with single charge (Supplementary S4).

Equifrequency curves at four frequencies are shown in Figure 2c-f. We plot equifrequency curve at 5.71 GHz where the type 2 Weyl points occur and the equifrequency curves of other frequencies are type I. The Weyl points marked as yellow circle are shown in Fig. 2e, whose Chern number is calculated by non-abelian Wilson loop methods.

This double helix structure with square lattice has a Brillouin zone of a cuboid. For clarification, all point and line degeneracies (including those not shown in Figure 2) between the lowest three bands and their coordinate $(k_x, k_y, k_z, f)$ are plotted in Figure 3 with the same notation in Figure 2.

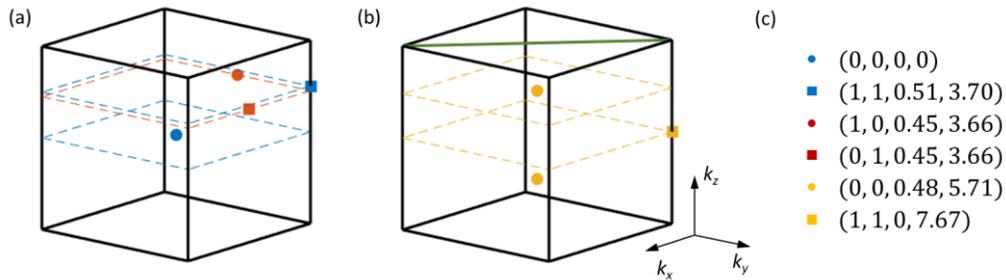

Figure 3. Brillouin zone with degeneracies between (a) the first and the second bands and (b) the second and the third bands. Circle and square points with the same color represent distinct Weyl points forming a pair. Blue circle point and yellow square point are doubly charged while the green line is topologically trivial. (c) $(k_x, k_y, k_z, f)$ at each point.

**Topologically protected surface states**

According to the bulk-edge correspondence, topological surface states are expected between the topologically inequivalent bands. Surface state dispersion is simulated by using supercell structure consisting of 19 unit cells aligned along x-axis, with boundary condition of perfect electric conductor, while the boundary conditions of the other directions are set to be periodic. Surface state dispersion at five different $k_z$ are shown in Figure 4a-e.

Between the first and the second bands, there exist topological surface states connecting the blue circle and square points. As one Weyl point is located at origin $(f = 0)$ and the other point at $f = 3.70$, the double helix structure supports topological surface states at all frequencies below 3.70 GHz. As opposed to previously reported topological surface states which have both lower and upper frequency limit, the surface states observed in the double helix structure has no lower limit and are hence extremely broad band. Other surface states connecting the red circle and square points are also found. Here, the surface states connecting the Weyl point pair marked as red points only appear in narrow bandwidth near 3.66 GHz where the point degeneracies are located. Figure 4 also shows the existence of topological surface states between the second and the third bands. As indicated in Figure 3d-e, the number of topological surface states varies as the frequency crosses the Weyl points. When $|k_z|$ is larger than 0.48 within which all Weyl points are located, the surface states disappear due to the neutralized topological charges.

Electric field distributions in logarithmic scale of two counter-propagating surface state at 3 GHz are shown in Figure 4g and h. Gradient of electric field amplitude indicates highly localized surface waves at each side. If the surface states are indeed topologically protected, then their propagation should be extremely robust against perturbations such as defects and cannot be back-scattered. To confirm this topological property, surface waves propagating along the sharp edge are shown in Figure 4i. Since the double helix structure is working in a metamaterial regime where unit cell is much smaller than the wavelength, homogeneous medium with the retrieved effective parameter (shown in supplementary S1) is used instead of the real structure. Line current excites specific $k_z$ between the topologically non-trivial band gap, generating robust surface waves which are not scattered by an arbitrarily shaped obstacle. Although we only focus on the surface states between three lowest bands, surface states connecting the higher bands are also observed (shown in Figure 4f), indicating the double helix structure works in photonic crystal regime as well.

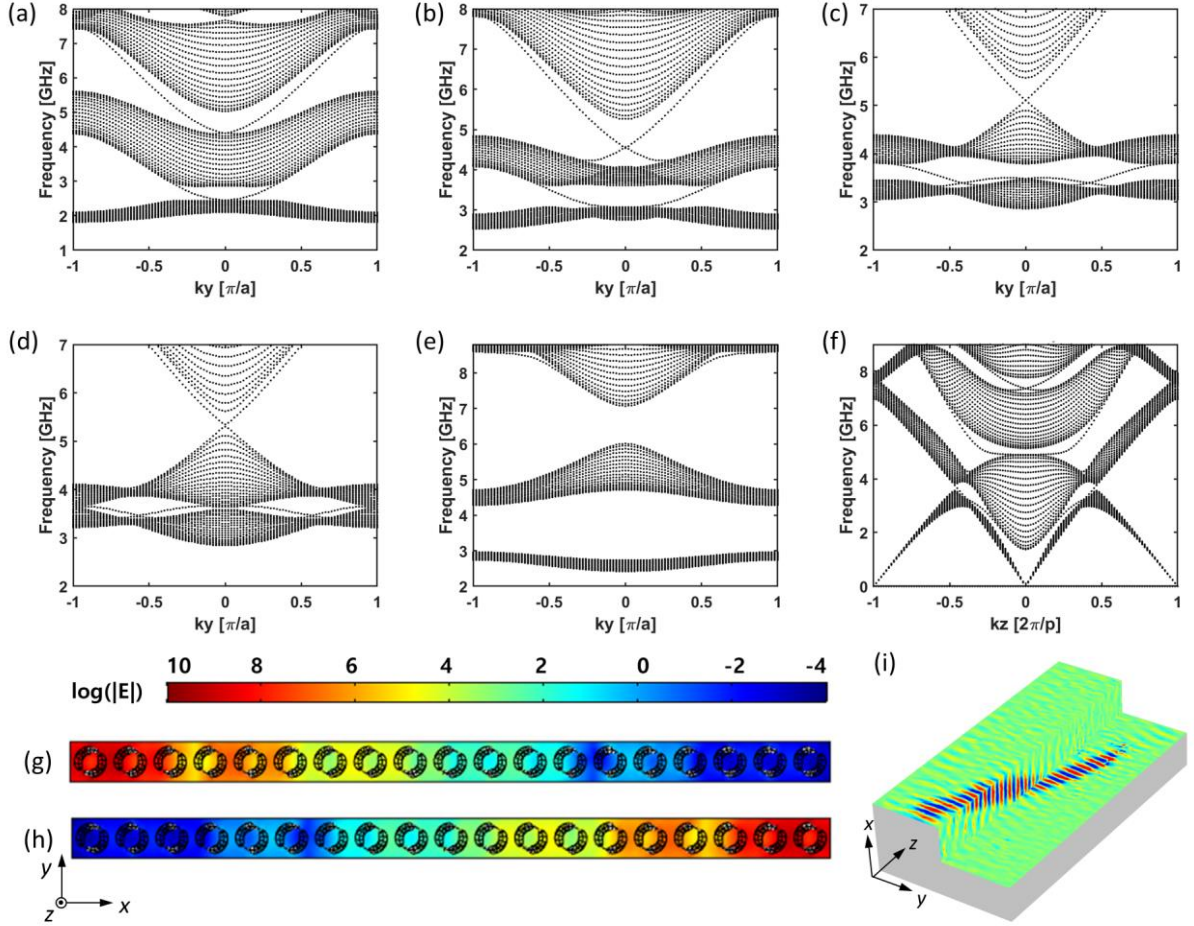

Figure 4. Band structures with surface states when $k_x = 0$ and (a) $k_z = 0.25$, (b) $k_z = 0.35$, (c) $k_z = 0.42$, (d) $k_z = 0.45$, (e) $k_z = 0.6$ and (f) $k_y = 0.25$. Electric field amplitude in logarithmic scale of surface state at 3 GHz when $k_x = 0$, $k_y = 0.25$, (g) $k_z = 0.317$ and (h) $k_z = -0.317$. (i) $x$-component of electric field of surface waves at 3 GHz.

**Discussion**

In conclusion, double Weyl points and extremely broadband topological surface states residing in a double helix structure are presented. Unlike topological photonic crystals, this topological metamaterial relies on effective hyperbolicity and chirality, which forces a Weyl point with double topological charge in zero frequency. We numerically observe six pairs of Weyl points and broadband topological surface states connecting them, which are robust against deformation. The extreme broad bandwidth and robustness will be beneficial for applications such as one-way waveguide and photonic integrated circuits.

**Acknowledgements**

This work was financially supported from the National Research Foundation of Korea (NRF) grants (NRF-2018M3D1A1058998, NRF-2015R1A5A1037668 and CAMM-2014M3A6B3063708) funded by the Ministry of Science and ICT (MSIT) of the Korean government. M.K. acknowledges Global Ph.D. Fellowships (NRF-2017H1A2A1043204) from NRF-MSIT of Korean government.

**Author contribution**

M.K. and J.R. initiated the research. M.K. designed the sample and conducted numerical simulation. M.K., D.L., T.H. and T.T.K. analyzed data. M.K. wrote the manuscript. J.R. and S.Z. guided the research. All authors participated in the discussion and approved the final manuscript.

**Competing interests**

The authors declare no competing interests.